\documentclass[12pt,preprint]{aastex}

\begin{document}

\title{Formation of Magnetically Supported Disks During Hard-to-Soft 
Transition in Black Hole Accretion Flows}

\author{Mami \textsc{Machida}}%
\affil{Division of Theoretical Astronomy, 
National Astronomical Observatory of Japan, \\
2--21--1 Osawa, Mitaka-shi, Tokyo 181--8588 }
\email{mami@th.nao.ac.jp}

\author{Kenji \textsc{Nakamura}}
\affil{Department of Sciences, Matsue National College of Technology, \\
14-4 Nishiikuma-cho, Matsue, Shimane 690-8518}
\email{nakamrkn@matsue-ct.jp}
\and
\author{Ryoji {\sc Matsumoto}}
\affil{Department of Physics, Faculty of Science, Chiba University, \\
 1-33 Yayoi-cho, Inage-ku, Chiba, 263-8522}
\email{matumoto@astro.s.chiba-u.ac.jp}

%



\begin{abstract}
We carried out three-dimensional global resistive
magnetohydrodynamic (MHD) simulations of the cooling instability in 
optically thin hot black hole accretion flows 
by assuming bremsstrahlung cooling.  
General relativistic effects are simulated by using the pseudo-Newtonian
potential. 
Cooling instability grows when the density of the accretion disk 
becomes sufficiently large. 
We found that as the instability grows the accretion flow 
changes from an optically thin, hot, gas pressure-supported 
state (low/hard state) to a cooler, 
magnetically supported, quasi-steady state. 
During this transition, magnetic pressure exceeds the gas pressure 
because the disk shrinks in the vertical direction almost conserving 
the toroidal magnetic flux. 
Since further vertical contraction of the disk is suppressed by 
magnetic pressure, the cool disk stays in an optically thin, 
spectrally hard state. 
In the magnetically supported disk, the heating rate balances 
with the radiative cooling rate. 
The magnetically supported disk exists for time scale 
much longer than the thermal time scale and  
comparable to the accretion time scale.

We examined the stability of the magnetically supported disk analytically, 
assuming that the toroidal magnetic flux is conserved, 
and found it thermally and secularly stable. 
Our findings may explain why black hole candidates stay in luminous, 
X-ray hard state 
even when their luminosity exceeds the threshold for the 
onset of the cooling instability. 
\end{abstract}


\section{Introduction}
State transitions between a low/hard state and a high/soft state 
are observed  in Galactic black hole candidates such as 
Cyg X-1 (e.g., \cite{hol75}; \cite{oda77}; \cite{zdz02}),  
GS 1124-68 (e.g., \cite{kit92}; \cite{ebi94}),   
GRS 1915+105 (e.g., \cite{bel00}; \cite{ued02})  
and GX 339-4 (e.g., \cite{miy91}, 1995; 
\cite{zdz04}; \cite{hom05}).  
Recent RXTE observations of state transitions in binary black hole 
candidates and their relation to jet ejections are reviewed by 
Fender, Belloni \& Gallo (2004) and \citet{rem05}.
The low/hard state is characterized by violent X-ray fluctuations and 
absence of the soft blackbody component in its spectrum.  
In this state, mass accretes to the black hole as an optically thin, 
advection dominated accretion flow 
(e.g., \cite{ich77}; \cite{nar94}, 1995).  
The spectrum of a high/soft state is dominated by blackbody-like 
component emitted from an optically thick, geometrically 
thin disk (e.g., \cite{sha73}). 
Recent X-ray observations of outbursts of GX 339-4 indicate that 
the source stays in the spectrally hard and luminous intermediate state 
during the transition from low/hard state to the high/soft state
(\cite{hom05}; \cite{rem05}). 
It has been suggested that relativistic ballistic ejections occur during 
the transition from this intermediate state to the optically thick, 
soft state \citep{fen04}. 

Outbursts of black hole candidates are triggered by the 
increase of the mass accretion rate from the outer radius 
\citep{min89}. 
In the early stage of an outburst, since a hot, optically thin, 
hard X-ray emitting disk is formed around the black hole, 
the black hole 
candidate is observed as staying in the low/hard state. 
The X-ray luminosity of the disk increases as the accretion rate increases.  
\citet{abr95} obtained thermal equilibrium curves of
optically thin accretion disks 
including both the bremsstrahlung cooling and advective cooling 
(see also \cite{kat98}). 
They showed that when the mass accretion rate (and, hence, surface density) 
exceeds some limit, 
optically thin disks cannot be in thermal equilibrium 
because the radiative cooling overwhelms the heating. 
When this limit is reached, therefore, a thermal instability is 
triggered so that the disk should cool and shrink in the 
vertical direction (\cite{min96}; Lasota, Narayan, \& Yi 1996). 

During this vertical contraction of the disk, 
magnetic fields in the turbulent disk will be amplified due to flux 
conservation or will be dissipated by magnetic reconnection. 
In the former case, strongly magnetized disk will be formed. 
The possible existence of such low-$\beta$ 
($\beta \equiv P_{\rm gas}/ P_{\rm mag}<1$) disk was pointed out 
by Shibata, Tajima \& Matsumoto (1990). 
Although magnetic flux can buoyantly escape from the disk by the 
Parker instability \citep{par66}, 
once low-$\beta$ disk is formed, the disk stays in the low-$\beta$ 
state because magnetic tension prevents the growth of the 
Parker instability. 
They called such disks as {\it magnetic cataclysmic disks} 
because explosive events will take place 
when the magnetic energy stored in the disk is released. 
Mineshige, Kusunose \& Matsumoto (1995) suggested that 
low-$\beta$ disks may correspond to the low/hard state. 
Recently Pariev, Blackman and Boldyrev (2003) constructed a model 
of an optically thick, low-$\beta$ disk based on 
$\alpha$-prescription of viscosity. 
However, they did not show how the low-$\beta$ disks are created. 
Three-dimensional resistive MHD simulations can answer this question. 

Global 3D MHD simulations of radiatively inefficient 
accretion disks have been carried out by several authors 
(e.g., \cite{mat99}; \cite{haw00},2001; 
\cite{mac00}; \cite{arm03}; \cite{mac03}; 
\cite{igm03}; \cite{kat04}).  
It has been shown that 
the MHD turbulence driven by magneto-rotational instability (MRI) 
enhances the angular momentum transport
rate and enables mass accretion. 
Initially weak magnetic fields are amplified due to the MRI, 
and saturates when the plasma-$\beta$ is about $10$.  
Machida, Nakamura \& Matsumoto (2004) showed that 
the radial structure of its innermost region 
($r < 20 r_g$, where $r_g$ is the Schwarzschild radius) 
agrees well with that of one-dimensional steady transonic solutions,  
in which the radial advection and the radial dependence of 
$\alpha$ are both taken into account. 
As the accretion rate $\dot{M}$ increases, 
the surface density of the disk $\Sigma$ increases. 
Numerical solutions follow the
evolution track $\Sigma \propto \dot{M}$
expected from conventional theory of optically thin, 
advection dominated hot accretion flows (e.g., \cite{abr95}).

The purpose of this paper is to study the 
growth of the thermal instability which takes place 
when the density of the optically thin disk becomes sufficiently large. 
We first simulated the formation of an optically thin, 
turbulent disk  without including the 
radiative cooling. 
Subsequently, we simulated the growth of the 
thermal instability by including the radiative cooling term. 

In section~\ref{method}, we describe basic equations, initial conditions, 
and numerical method. 
The results of simulations are given in section~\ref{result}.  
Section~\ref{discuss} is devoted for discussion and conclusion.

\section{Numerical Model} \label{method}

\subsection{Basic Equations}

We solved the following resistive MHD equations in a cylindrical 
coordinate system $(\varpi, \varphi, z)$; 

%
%
\begin{equation}
       \frac{\partial \rho}{\partial t} 
    +  \nabla ( \rho \mbox{\boldmath $v$} )
  = 
       0 ~,
\label{eqn:b1}
\end{equation}
%
%
\begin{equation}
      \rho \left[
      \frac{\partial \mbox{\boldmath $v$}}{\partial t}
    + \mbox{\boldmath $v$} \cdot \nabla \mbox{\boldmath $v$}
      \right]
  = 
    - \nabla P 
    - \rho \nabla \phi
    + \frac{\mbox{\boldmath $j$} \times \mbox{\boldmath $B$}}{c} ~,
\label{eqn:b2}
\end{equation}
%
%
\begin{equation}
     \frac{\partial \mbox{\boldmath $B$}}{\partial t}
  = 
     \nabla \times 
    ( \mbox{\boldmath $v$} \times \mbox{\boldmath $B$} 
    - \frac{4 \pi}{c} \eta \mbox{\boldmath $j$} ) ~,
\label{eqn:b3}
\end{equation}
%
%
\begin{equation}
     \rho T \frac{d S}{dt}
  = 
    \frac{4 \pi}{c^2} \eta j^2 - Q_{\rm rad}^{-} ~,
\label{eqn:b4}
\end{equation}
where $\rho$, $P$, $\phi$, $\mbox{\boldmath $v$}$, $\mbox{\boldmath $B$}$, 
$\mbox{\boldmath $j$} = c \nabla \times \mbox{\boldmath $B$}/4 \pi$, 
$\eta$, $T$ and $S$ are the density, pressure, 
gravitational potential, velocity, magnetic field, current density, 
resistivity, temperature and specific entropy, respectively. 
The specific entropy is expressed as 
$S = C_{\rm v} {\rm ln} {(P/ \rho^{\gamma})}$ 
where $C_{\rm v}$ is the specific heat capacity and 
$\gamma$ is the specific heat ratio. 
The Joule heating term and the radiative cooling term 
$Q_{\rm rad}^{-}$ are included in the energy equation. 
We assume the anomalous resistivity 
$\eta = \eta_0 [{\rm max} (v_{\rm d}/v_{\rm c}-1,0)]^2$ 
\citep{yok98} 
where $v_{\rm d} \equiv j/\rho$ is the electron-ion drift speed 
and $v_{\rm c}$ is the threshold above which 
anomalous resistivity sets in. 
We assume bremsstrahlung cooling 
$Q_{\rm rad}^- = 6.2 \times 10^{20} \rho^2 T^{1/2} ~ 
{\rm erg} ~ {\rm s}^{-1} ~ {\rm cm}^{-3}$.  

General relativistic effects are simulated by using the 
pseudo-Newtonian potential $\phi = -GM/(r-r_g)$ \citep{pac80}, 
where $G$ is the gravitational constant, $M$ is the mass of the 
black hole, and $r=(\varpi^2 + z^2)^{1/2}$.
We neglect the self gravity of the disk.

\subsection{Numerical Methods and Boundary Conditions}

We solved the resistive MHD equations in a cylindrical 
coordinate system $(\varpi,
\varphi, z)$ by using a modified Lax-Wendroff scheme \citep{rub67} with an
artificial viscosity \citep{ric67}.  

For normalization, we use units listed in table \ref{tab1}.  
The units of length and velocity are Schwarzschild radius $r_g$ 
and the light speed $c$. 
For unit density, we adopt $\rho_0 = 1/(\kappa_{\rm es} r_g)
= 8.3 \times 10^{-7} (M/10M_{\odot})~ {\rm g} ~ {\rm cm}^{-3}$, 
where $\kappa_{\rm es} = 0.4 ~ {\rm cm}^2/{\rm g}$ 
is the electron scattering opacity. 
The unit time is $t_0 = r_g/c = 10^{-4} (M/10M_{\odot})~ {\rm sec}$. 
The unit temperature is given by 
$T_0 = m_p c^2/k_{\rm B}= 1.1 \times 10^{13} ~ {\rm K}$, 
where $m_p$ is the proton mass and $k_{\rm B}$ is the Boltzmann constant. 
The accretion rate is normalized by 
$\dot{M}_{\rm Edd} = L_{\rm Edd}/c^2 $, 
where $L_{\rm Edd} = 4 \pi c GM/ \kappa_{\rm es}
= 1.25 \times 10^{39} (M/10 M_{\odot})$  ${\rm erg ~ s^{-1}}$ is the 
Eddington luminosity. 

In the energy equation, we use the normalized cooling rate 
\begin{equation}
   Q_{\rm rad}^{'} 
 = 
   \frac{Q_{\rm rad}^-}{(\rho_0 c^3/ r_g)} 
 = 
   Q_{\rm b}^{'} 
   \left( \frac{\rho}{\rho_0} \right)^2 
   \left( \frac{T}{T_0} \right)^{1/2} ~.
\end{equation}
Here, $T$ is ion temperature and 
$Q_{\rm b}^{'}$ is a parameter depending on the electron-ion 
coupling and inverse Compton amplification. 
In this paper, we assume single temperature plasma and 
adopt $Q_{\rm b}^{'} = 1.9 \times 10^{-4}$.

We solved the energy equation transformed to the conservation form; 
\begin{equation}
\frac{\partial}{\partial t} \left(
 \frac{1}{2}\rho^{'} v^{' \, 2} + \frac{B^{' \, 2}}{8 \pi} + 
 \frac{P^{'}}{\gamma-1} \right)
+ \nabla \cdot \left[ \left(
 \frac{1}{2} \rho^{'} \mbox{\boldmath $v^{'}$}^2 + 
 \frac{\gamma P^{'}}{\gamma-1} 
 \right) \mbox{\boldmath $v^{'}$} + 
 \frac{1}{4 \pi} \mbox{\boldmath $E^{'}$} \times 
 \mbox{\boldmath $B^{'}$} \right]
 = - \rho \mbox{\boldmath $v^{'}$} \nabla \phi^{'}
   - Q_{\rm rad}^{'}
\end{equation}
where $\mbox{\boldmath $E^{'}$} = - \mbox{\boldmath $v^{'}$} \times 
\mbox{\boldmath $B^{'}$} + 4 \pi \eta \mbox{\boldmath $j^{'}$}$ 
and $'$ denotes the normalized quantities. 
In the following, we omit $'$ from the symbols for 
simplicity except the radiative cooling term.  
In this formulation of the energy equation, 
the numerically dissipated kinetic energy and magnetic energy are 
captured as the thermal energy \citep{hir05}. 
Shock heating is also incorporated self-consistently. 
When magnetic turbulence develops in the disk, 
the disk plasma is heated more efficiently than expected from 
the Joule heating. 
However, in the early stage of the simulation when the disk 
is still laminar, since we did not include the explicit 
heating term except the Joule heating, the disk heating is 
insufficient to balance with the cooling term.
Thus, we include the radiative cooling term after a quasi-steady 
turbulent accretion disk is formed through accretion from the initial torus.

The number of grids is $(N_{\varpi}, N_{\varphi}, N_z) = (250, 64, 191)$.  
The grid size is $\Delta \varpi = \Delta z = 0.1 $ for 
$0 < \varpi, ~ z < 10 $, and otherwise increases with $\varpi$ 
and $z$. 
The grid size in azimuthal direction is $\Delta \varphi = 2 \pi / 63$. 
We simulated only the upper half space $(z \geq 0)$ by applying 
a symmetric boundary condition at the equatorial plane $(z = 0)$. 
The outer boundaries at $\varpi = 132$ and 
at $z =69$ are free boundaries where waves can be transmitted.  
We included the full circle $(0 \leq \varphi \leq 2 \pi)$ 
in the simulation region, and 
applied periodic boundary conditions in the azimuthal direction. 
An absorbing boundary condition is imposed  at 
$r = r_{\rm in} = 2 $ 
by introducing a damping parameter; 
\begin{equation}
       a
  =  
       0.1 \left(
       1.0 - \tanh{
       \frac{r - r_{\rm in} + 5 \Delta \varpi}{2 \Delta \varpi}}
       \right) ~.
\label{eqn:b6}
\end{equation}
Physical quantities $q = (\rho, \mbox{\boldmath $v$}, \mbox{\boldmath $B$}, 
P)$ inside $r = r_{\rm in}$ are re-evaluated by 
\begin{equation}
       q^{\rm new} 
 = 
       q - a(q - q_0) ~,
\label{eqn:b7}
\end{equation}
which means that the deviation from initial values $q_0$ is 
artificially reduced with damping rate $a$. 
Waves propagating inside $r=r_{\rm in}$ are absorbed in the transition 
region ($r_{\rm in}-5 \Delta \varpi < r < r_{\rm in}$).

\subsection{Initial Model}

The initial state of our simulation is an equilibrium torus threaded by weak
toroidal magnetic field \citep{oka89}. 
Initially the torus is assumed to have a constant specific 
angular momentum $L$.

According to \citet{oka89},  
by using the polytropic relation $P=K \rho^{\gamma}$ 
at the initial state and by assuming 
\begin{equation} 
 \beta = \frac{8 \pi P}{B_{\varphi}^2} 
       = \beta_{\rm b} \left( \frac{\varpi}{\varpi_b}
       \right)^{-2(\gamma-1)}
\end{equation}
where $\beta_{\rm b}$ is the initial plasma $\beta$ at the 
initial pressure maximum of the torus 
$(\varpi,z) = (\varpi_{\rm b},0)$, and $B_{\varphi}$ is the 
azimuthal magnetic field, 
we integrated the equation of motion into a potential form;
\begin{equation}
      \Psi(\varpi,z) 
 = 
      \phi + \frac{L^2}{2 \varpi^2} + \frac{1}{\gamma-1}v_{\rm s}^2 
     + \frac{\gamma}{2(\gamma -1)}v_{\rm A}^2 = \Psi_{\rm b}  
 =  {\rm constant} ~ , 
\label{eqn:poten}
\end{equation}
where $v_{\rm s} = (\gamma P/\rho)^{1/2}$ is the sound speed, 
$v_{\rm A} = [B_{\varphi}^2/(4 \pi \rho)]^{1/2}$ is the Alfv{\'e}n speed, 
and $\Psi_{\rm b} = \Psi(\varpi_{\rm b},0)$.  
At $\varpi = \varpi_{\rm b}$, the rotation speed of the torus 
$L/\varpi_{\rm b}$ equals the Keplerian velocity. 
By using equation (\ref{eqn:poten}), we obtain the density distribution as 
\begin{equation}
      \rho 
 = 
      \rho_{\rm b} \left(
         \frac{\max{ \{\Psi_{\rm b} - \phi -L^2/(2\varpi^2),0\}}}
              {K[\gamma/(\gamma-1)]
                [1+\beta_{\rm b}^{-1}\varpi^{2(\gamma-1)}
       /\varpi_{\rm b}^{2(\gamma-1)}]} 
       \right)^{1/(\gamma -1)}      ~ ,   
\end{equation}
where $\rho_{\rm b}$ is the density at 
$(\varpi,z) = (\varpi_{\rm b},0)$. 
Outside the torus, we assumed a hot, isothermal ($T=T_{\rm halo}$) 
spherical halo. 
The density distribution of the halo is given by 
$\rho_{\rm h} = \rho_{\rm halo} 
\exp [-(\phi-\phi_{\rm b})/(k_{\rm B} T_{\rm halo})]$, 
where $\phi_{\rm b}$ is the gravitational potential at 
$(\varpi,z)=(\varpi_{\rm b},0)$. 
We assume that the radiative cooling is negligible in the coronal region 
where the density is lower than $\rho_{\rm crit} = 10^{-4} \rho_{\rm b}$.

In this paper, we adopted model parameters $\varpi_{\rm b} = 50$, 
$K = P_{\rm b}/ \rho_{\rm b}^{\gamma} = 5 \times 10^{-4}$, 
$\beta_{\rm b} = 100$, $\gamma = 5/3$, $L=(\varpi_{\rm b}/2)^{1/2}
\varpi_{\rm b} /(\varpi_{\rm b}-1)$, 
$\rho_{\rm halo} = 10^{-4} \rho_{\rm b}$, 
$\eta_0 = 5 \times 10^{-4}$, and $v_{\rm c}= 0.9$. 
The inner and outer radii of the initial torus are 
$\varpi_{\rm in}=33$, and $\varpi_{\rm out}=85$, respectively.

\section{Numerical Results} \label{result}

\subsection{Evolution of a Thermally Unstable Accretion Disk}

\begin{figure}
  \epsscale{1.0}
  \plotone{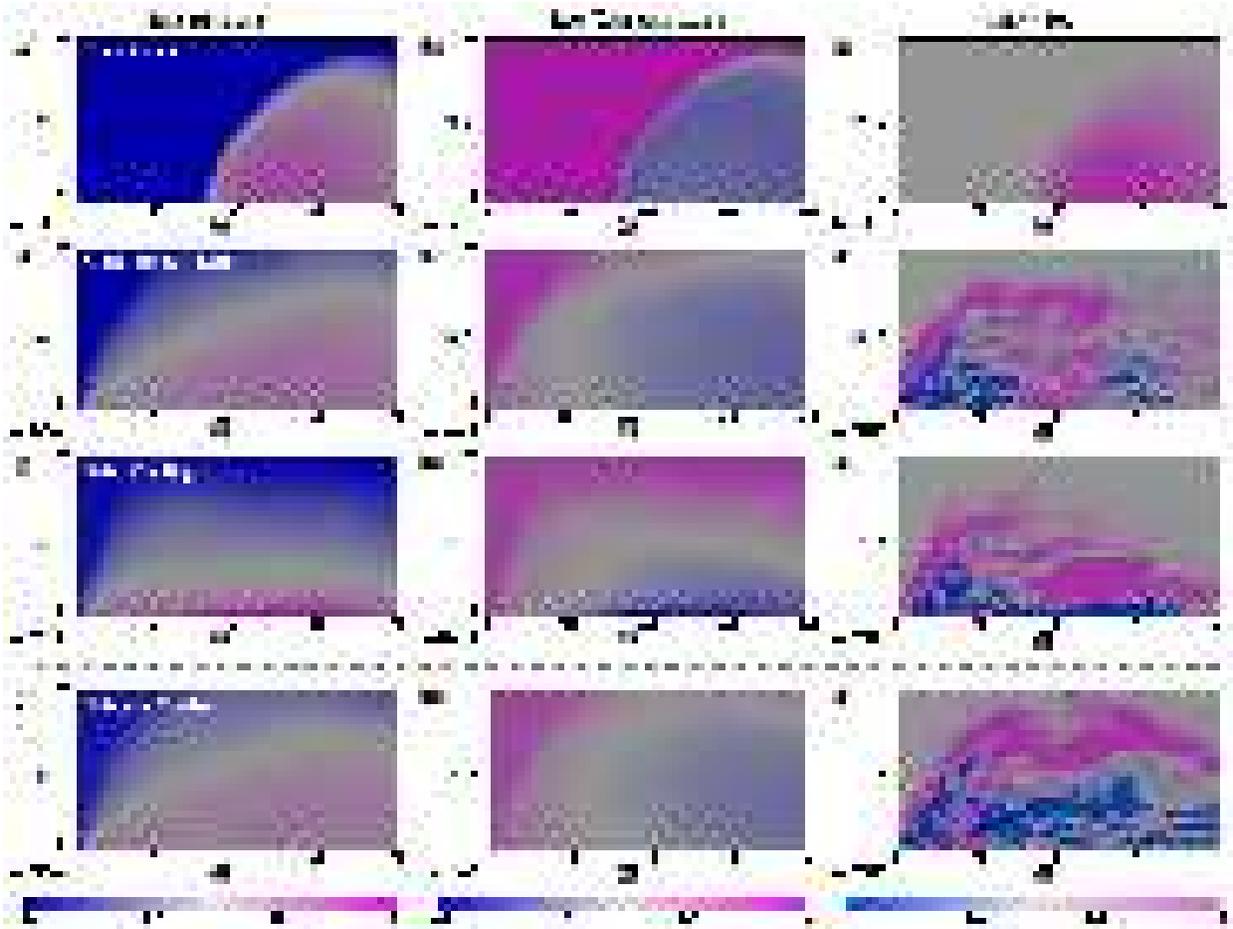}
  \caption{
Distribution of density (left), temperature (middle) and 
azimuthal magnetic field (right) averaged in azimuthal direction. 
Top panels show the initial state. 
Second panels from top show the stage before the transition ($t=24000$). 
Third panels from top show the stage after cooling instability sets in
($t=32000$). 
Bottom panels show the same stage as the third panels ($t=32000$) 
for a model without including radiative cooling effect. }
  \label{fig:fig2}
\end{figure}

Here we present the results of the numerical simulation for 
a model with $\rho_{\rm b} = 0.29$. 
Figure \ref{fig:fig2} shows the time evolution of azimuthally 
averaged density $\log \langle \rho / \rho_{\rm b} \rangle$, 
temperature and azimuthal component of magnetic fields.  
The top panels show the initial state. 
Figure \ref{fig:fig2}d, \ref{fig:fig2}e and \ref{fig:fig2}f show the stage 
just before the cooling term is switched on. 
The initial torus deforms itself into an accretion disk 
by transporting angular momentum through Maxwell stress. 
Azimuthal magnetic fields are not completely random but show 
sector structures. 

We switched on the radiative cooling term at $t=24100$ 
when the disk becomes quasi-steady. 
At this stage, the Thomson optical thickness of the disk is 
$\tau_{\rm d} = \kappa_{\rm es} \Sigma \sim 3$.  
Here, the surface density $\Sigma$ is computed by vertically 
integrating the azimuthally averaged density 
from $z = - H_{\rm d}$ to $z = H_{\rm d}$, 
where $H_{\rm d}$ is the half thickness of the disk, 
which we take $H_{\rm d} = 0.3 \varpi + 13.5$. 
The disk is thick for Thomson scattering 
($\tau_{\rm d} \sim 3$) but effectively optically thin 
(see e.g., \cite{kat98}). 
This state corresponds to the low/hard state observed in  
black hole candidates. 

The third panels from top in figure \ref{fig:fig2} show the stage after 
the onset of the cooling instability. 
Geometrically thin, cool, dense disk is created in 
$15 < \varpi < 70$. 
Magnetic fields in the equatorial region of the vertically shrinking 
cool disk become almost toroidal and well-ordered. 
In the inner region ($\varpi < 15$), 
the disk still stays in the hot state.

For comparison, in the bottom panels in figure \ref{fig:fig2}, 
we show the numerical results at $t = 32000$ for a 
model without including the radiative cooling term. 
The disk continues to stay in the hot, 
magnetically turbulent state. 
Between $t=24000$ and $t=32000$, 
the disk structure does not change significantly. 

We also confirmed that even when we included the cooling term, 
if the initial density of the torus is small enough 
($\rho_{\rm b} \le 0.05$), 
the whole disk stays in optically thin hot state 
throughout the simulation $(0 \le t \le 32000)$.

\begin{figure}
  \epsscale{0.8}
  \plotone{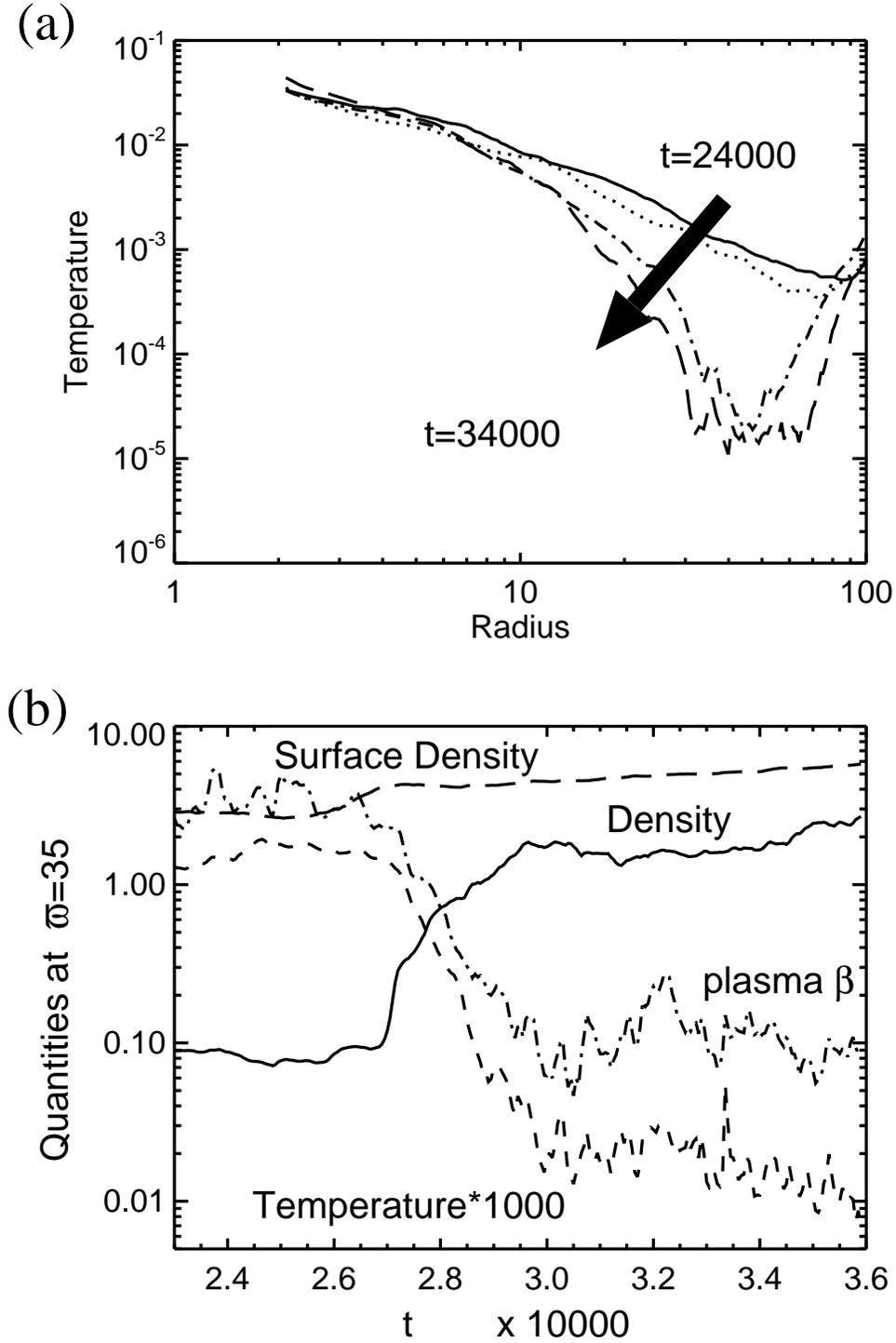}
  \caption{
(a) Radial distribution of temperature averaged in azimuthal 
direction and in vertical direction $(0 < z <1)$. 
Curves show the temperature at 
$t= 24000$, $26000$, $30000$, and $34000$ 
from top to bottom.
(b) Time evolution of azimuthally averaged surface density, 
equatorial density, equatorial temperature, and equatorial 
plasma $\beta$ at $\varpi = 35$. 
The units of the horizontal axis is $t= 10^4$  
[$= 1.0 (M/10M_{\odot})$ sec in physical units].
}
  \label{fig:fig3}
\end{figure}

Figure \ref{fig:fig3}a shows the radial distribution of the temperature 
averaged in the azimuthal direction and in the vertical direction 
($0 < z <1$). 
Curves show the temperature at $t=24000$, $26000$, 
$30000$, and $34000$ from top to bottom. 
This figure indicates that cooling front 
propagates inward. 
The temperature in the cool region decreases down to 
$10^{-5}$ ($ \sim 10^8 {\rm K}$ in physical units). 
The region $\varpi < 15$ stays in the hot state 
throughout the simulation. 
The inner region stays in the hot state 
because the accretion rate is not large enough to trigger the 
thermal instability in this region. 
Note that in conventional models of optically thin, hot 
accretion flows including bremsstrahlung cooling, 
the critical accretion rate for the onset of the cooling 
instability increases with decreasing the radius
(Figure 2 in \cite{abr95}).

Figure \ref{fig:fig3}b shows the time evolution of azimuthally averaged 
surface density, equatorial density, equatorial temperature and 
equatorial plasma $\beta$ at $\varpi = 35$. 
Here, surface density is integrated in $|z| < 24$. 
Numerical results indicate that after the inclusion of radiative 
cooling, the disk evolves quasi-steadily until $t=27000$. 
Subsequently, the disk shrinks in the vertical direction due to 
the cooling instability.  
As the cooling instability grows, plasma $\beta$ decreases 
and the disk becomes dominated by magnetic pressure 
(i.e. $\beta < 1$).

Let us compare the cooling time scale 
$t_{\rm cool} = P/ Q_{\rm rad}^{-} = (T/T_0)^{1/2} / 
[Q_{\rm b}^{'} (\rho/ \rho_0)]$ with other time scales. 
Before the onset of the cooling instability, 
since $T \sim 10^{-3} $ [$ \sim 10^{10} {\rm K}$ in physical units]
and $\rho \sim 0.1 $ [$\sim 8.3 \times 10^{-8} 
(M/10 M_{\odot})$  g ${\rm cm}^{-3}$], 
the cooling time scale is  
$t_{\rm cool} \sim 1.6 \times 10^{3}$  
[$\sim 0.16 (M/10M_{\odot})$ sec in physical units]
at $\varpi=35$. 
This time scale is comparable to the rotation time scale 
$\sim 2 \pi t_{\rm dyn} \sim 1.8 \times 10^3 $ 
where $t_{\rm dyn} = 1/ \Omega \sim 300 $ is the dynamical 
time scale at $\varpi  = 35$. 
Here $\Omega$ is the rotation angular speed.
The accretion time scale 
$t_{\rm acc} = \varpi/v_{\varpi} \sim 10^4 $ is an order of magnitude 
larger than the thermal time scale. 
Numerical results indicate that the temperature of the disk 
decreases from $T \sim 10^{-3}$ to $ T \sim 10^{-5} $ 
in $t \sim 6 \times 10^3$. 
This time scale ($\sim 4 t_{\rm cool}$) is consistent with 
the time scale for the growth of the thermal instability.

\begin{figure}
 \epsscale{0.5}
 \plotone{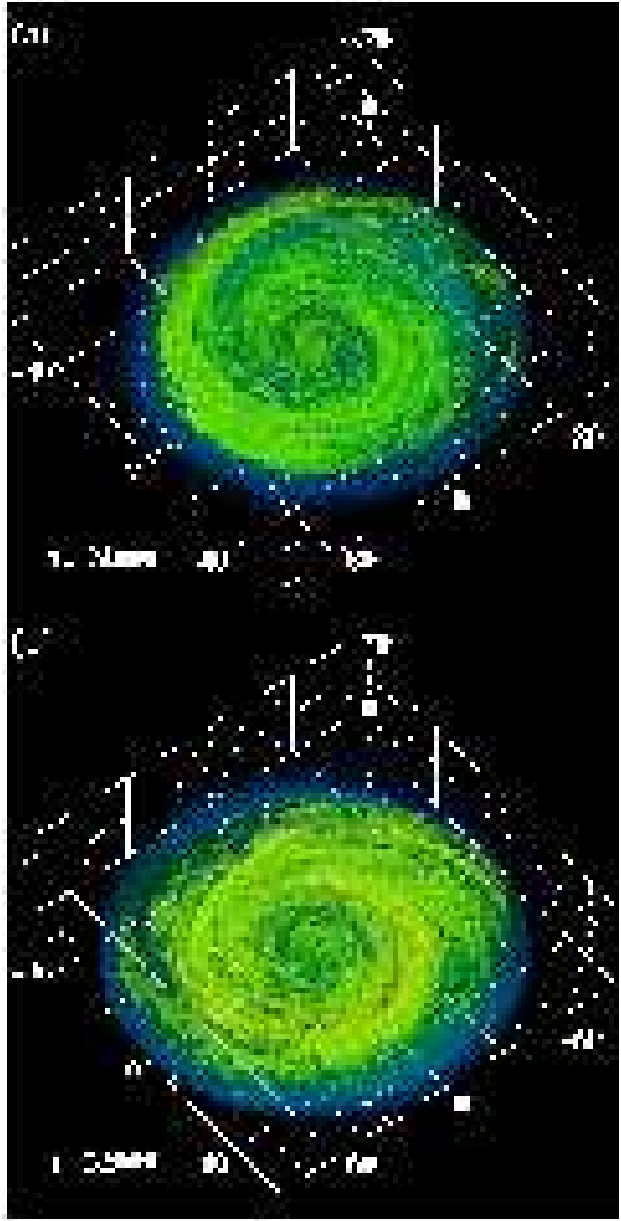}
 \caption{Isosurface of $\beta = P_{\rm gas}/ P_{\rm mag}$. 
The blue, green, and yellow surfaces correspond to $\beta=100$, 
$\beta=10$, and $\beta=1$, respectively.
(a) Before the transition, (b) during the transition.}
 \label{fig:beta}
\end{figure}

Figure \ref{fig:beta} shows the isosurface of the plasma $\beta$ at 
$t=24000$ and $t=32000$. 
The blue, green, and yellow surfaces show $\beta =100$, $10$, and $1$, 
respectively. 
Before the cooling term is switched on, low-$\beta$ regions 
($\beta < 1$) occupy only a small fraction of the disk. 
However, after the onset of the transition, 
the low-$\beta$ regions fill the disk. 
The filling factor of the low-$\beta$ region in $ |z| < 1$ 
increases from $0.1$ to $0.7$ during the transition. 

\subsection{Structure of Global Magnetic Fields}

\begin{figure}
  \epsscale{1.0} 
  \plotone{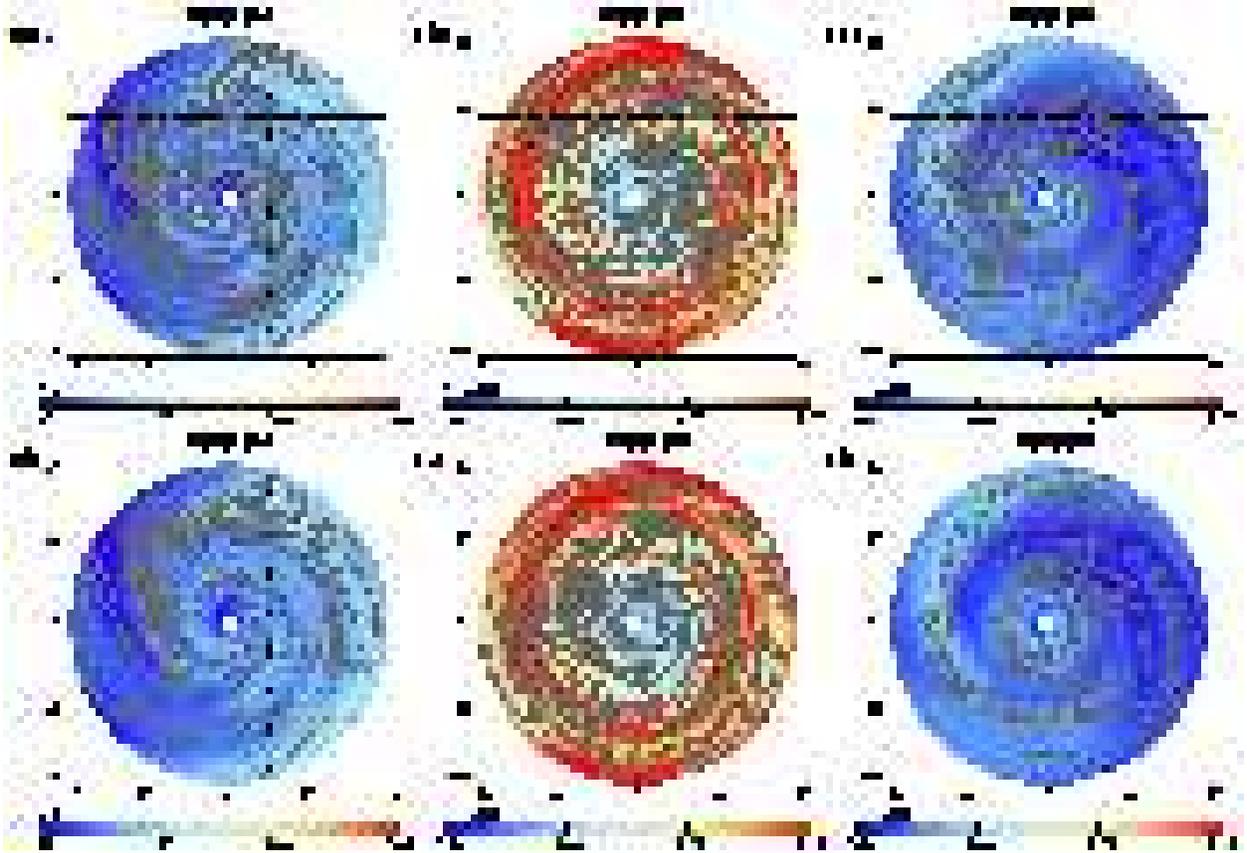}
  \caption{
Distribution of equatorial density (color) and magnetic field lines 
projected onto the equatorial plane (gray curves). 
(top) magnetic field lines depicted by original magnetic fields 
before decomposing them into the mean field and fluctuating field. 
(bottom) magnetic field lines depicted by mean magnetic fields.
(left) Before the cooling instability grows ($t=26000$). 
(center) After the cooling instability sets in ($t=32000$). 
(right) Result of a model without cooling ($t=32000$). }
  \label{fig:str}
\end{figure}

Top panels of figure \ref{fig:str} show the 
equatorial logarithmic density 
(color) and magnetic field lines projected onto 
the equatorial plane (gray curves). 
Before the cooling instability grows 
(figure \ref{fig:str}a), 
magnetic field lines show loosely wound global spiral structure 
superposed by turbulent components. 
As the disk cools, magnetic fields turn into a tightly 
wound spiral 
(figure \ref{fig:str}b).  
The equatorial density increases in the outer region 
($\varpi > 15$), where cooling instability grows. 
Figure \ref{fig:str}c shows the equatorial density and 
magnetic field lines at $t=32000$ for a model 
without radiative cooling. 
In this model, the equatorial density remains small and 
the magnetic fields show large amplitude fluctuations.

In low-$\beta$ disks, turbulent magnetic fields become 
smaller than those in hot, gas pressure dominated disks 
because magnetic tension suppresses the motion with short  
azimuthal wave length. 
In order to show this more clearly,  
we decomposed magnetic fields to mean magnetic field  
$\mbox{\boldmath ${\bar B}$}$ and turbulent magnetic field 
$\delta \mbox{\boldmath $B$} = \mbox{\boldmath $B$} - 
\mbox{\boldmath ${\bar B}$}$
where the mean magnetic field  $\mbox{\boldmath ${\bar B}$}$ 
is computed by averaging the magnetic field in the region 
$\pm 10$ grids in $\varpi$ and $z$ directions and $\pm 3$  
grids in azimuthal direction. 
The bottom panels in figure \ref{fig:str} show the 
equatorial density and magnetic field lines depicted by 
the mean magnetic fields $\mbox{\boldmath ${\bar B}$}$. 
Figure \ref{fig:str}e shows that as the disk cools, 
magnetic fields turn into a tightly wound spiral 
with well ordered mean magnetic fields.

\begin{figure}
 \epsscale{0.8}
 \plotone{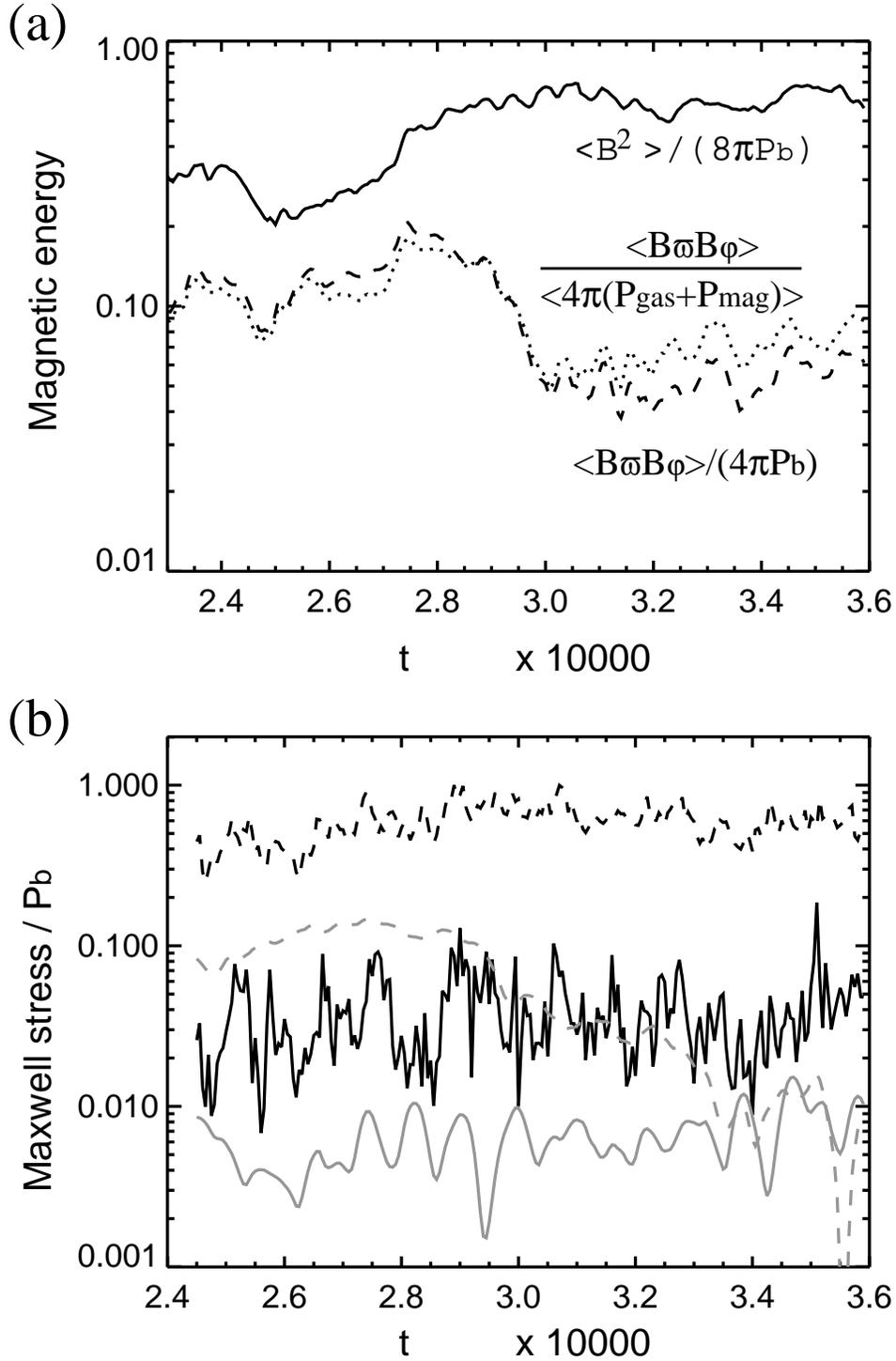}
 \caption{
(a) Time evolution of the azimuthally averaged magnetic pressure 
(solid curve) and the Maxwell stress (dashed curve) at $\varpi = 35$ 
normalized by the initial gas pressure and averaged in 
$0 < \varphi < 2 \pi$ and $0 < z < 1$.  
Dotted curve shows the ratio of the averaged Maxwell stress  
to the averaged total pressure $P_{\rm gas}+P_{\rm mag}$. 
(b) Time evolution of Maxwell stress computed by mean magnetic 
fields (solid curves) and fluctuating magnetic fields 
(dashed curves). 
The black curves show the Maxwell stress averaged in the 
inner region ($4 < \varpi < 10$ and $0 <z <1$). 
The gray curves show the Maxwell stress averaged in the outer region 
($30 < \varpi < 40$ and $0 <z <1$).}
 \label{fig:bxby}
\end{figure}

Figure \ref{fig:bxby}a shows the time evolution of the magnetic pressure 
and Maxwell stress at $\varpi=35$. 
The solid and dashed curves show the magnetic pressure and Maxwell 
stress normalized by the initial gas pressure averaged 
in the azimuthal direction and in the vertical direction ($0 <z < 1$). 
The magnetic pressure increases as the disk shrinks in the 
vertical direction by cooling instability. 
The average ratio of Maxwell stress to the total pressure 
$\alpha_{\rm B} = \langle B_{\varpi} B_{\varphi} /4 \pi\rangle / 
\langle P_{\rm gas} + P_{\rm mag} \rangle$
slightly decreases but stays around 
$\alpha_{\rm B} \sim 0.1$.

Although the magnetic energy increases when low-$\beta$ disk 
is formed, Maxwell stress does not increase because the 
fluctuating radial magnetic field decreases. 
Inside the low-$\beta$ disk, azimuthal component of magnetic 
field dominates the fluctuating radial magnetic field. 
Figure \ref{fig:bxby}b shows the time evolution of the 
Maxwell stress normalized by the initial gas pressure, 
$\langle B_{\varpi} B_{\varphi}/4 \pi \rangle
/ P_{\rm b}$ averaged in azimuthal direction, 
in radial direction, and in vertical direction
$(0 \le z \le 1)$.  
Black curves show the Maxwell stress averaged in the inner region 
($ 4 < \varpi < 10$), 
and gray curves show that averaged in 
the outer region ($30 < \varpi < 40$).
The solid curves and the dashed curves show the time variation of 
the Maxwell stress computed from the mean magnetic field 
$\langle {\bar B_{\varpi}} {\bar B_{\varphi}} 
/4 \pi \rangle / P_{\rm b}$ 
and that computed from the fluctuating magnetic field 
$\langle \delta B_{\varpi}
\delta B_{\varphi}/4 \pi \rangle / P_{\rm b}$. 
In the inner region where the disk still stays in the 
hot state, the angular momentum transport is 
mainly due to the fluctuating component. 
In the outer region, the angular momentum transport  
by fluctuating field decreases with time 
and becomes comparable to that by mean magnetic fields. 
This figure indicates that the magnetic turbulence is 
suppressed inside the magnetically supported disk 
but the angular momentum is still transported by the 
Maxwell stress of mean magnetic fields.

\subsection{Increase of Luminosity}

Figure \ref{fig:lum} shows the time variation of the X-ray luminosity 
computed by integrating  the optically thin bremsstrahlung cooling term 
in $ 4 < \varpi < 50$, $ |z| < 30$ (solid curve) 
and in $4 < \varpi < 50$, $|z| < 1$ (dashed curve) 
when $\rho_{\rm b} = 0.29$. 
This luminosity is normalized by the Eddington luminosity. 
X-ray luminosity increases during the growth of the thermal instability  
because the equatorial density increases 10 times while the temperature 
decreases 100 times. 
Thus the cooling rate in $|z| < 1$ becomes 10 times higher than 
that before the transition. 
In this stage, the effective optical depth in the low-$\beta$ region 
is about $0.01$. 
Thus the disk still stays in an optically thin state. 
The increase of the X-ray luminosity saturates because 
the mass accretion rate from the outer region saturates. 
In our simulations, the mass accretion rate in $30000<t<36000$ 
is $\dot{M} \sim \dot{M}_{\rm Edd}$ when $\rho_{\rm b} = 0.29$. 
Since the energy conversion rate $\eta_{e}$ from the rest mass energy 
to radiation is $\eta_e < 0.06$ 
in accretion disks around a Schwarzschild black 
hole, $ L/L_{\rm Edd} < 0.06 (\dot{M}/\dot{M}_{\rm Edd}) \sim 0.06$ 
when $\dot{M} \sim \dot{M}_{\rm Edd}$. 
The disk luminosity shown in figure \ref{fig:lum} is an order of 
magnitude smaller than this limit because the innermost region 
($\varpi < 15$) still stays in radiatively inefficient optically 
thin, hot state. 
The saturation level of the luminosity will increase 
if we start simulation with larger $\rho_{\rm b}$ because the 
maximum accretion rate from the outer torus increases as the  
initial density of the torus increases. 
When the accretion rate further increases, 
the conversion rate from rest mass energy to radiation will 
also increase because the inner region of the disk 
undergoes a transition to radiatively efficient cool disk. 
In this paper, we only included the bremsstrahlung cooling. 
It should be noted that in optically thin, hot, magnetized 
disks, synchrotron Compton cooling may also contribute 
to increase the X-ray luminosity.

\section{Discussion} \label{discuss}

\subsection{Evolution of Optically Thin Accretion Disks}

In this paper, we have shown by direct numerical simulations without 
assuming $\alpha$-viscosity 
that magnetically supported disk is created during the transition 
from a low/hard state to a high/soft state 
in black hole accretion flows.

Figure \ref{fig:f8} schematically shows the evolution of an 
optically thin accretion disk. 
As the accretion rate from the outer disk increases 
(figure \ref{fig:f8}a), the density of the inner accretion disk 
increases. 
When the density of the disk exceeds the critical density for 
the onset of the cooling instability at some radius, 
the disk cools and thus shrinks in the vertical direction 
(figure \ref{fig:f8}b). 
In such regions, magnetic pressure dominates the gas pressure 
because 
(1) gas pressure decreases due to cooling, while 
(2) the azimuthal magnetic field increases as the disk shrinks 
in the vertical direction. 
The latter is due to the formation of sector structures 
in azimuthal magnetic fields. 
If the magnetic fields in turbulent accretion disks are purely random, 
they will be dissipated by magnetic reconnection, when the 
disk shrinks in the vertical direction. 
If magnetic fields have coherent structures such that 
the azimuthal component has the same sign in some regions, 
however,  
they can survive during the contraction due to cooling. 
Moreover, the field strength increases as the region shrinks 
conserving the toroidal magnetic flux.

Since magnetic pressure supports the disk even when 
temperature decreases, the disk stops shrinking in the 
vertical direction and approaches a new equilibrium state. 
The magnetically supported disk emits power-law hard X-rays 
because the disk still stays in optically thin state. 
Since the critical surface density for the onset of the 
cooling instability in the inner radius is larger than that 
in the outer radius when bremsstrahlung cooling is assumed 
\citep{abr95}, the inward propagation of the cooling front 
stops and the innermost region stays in the hot state. 
We expect that when the accretion rate from the outer disk 
further increases, 
the transition radius between the hot, gas pressure dominated 
region and the cool, magnetic pressure dominated region will 
move inward (figure \ref{fig:f8}c).

The evolution of the innermost region of optically thin disks 
will depend on the cooling mechanism. 
When the synchrotron and/or Compton cooling were included, 
transition to the magnetically supported disk may start from 
the inner region because the cooling rate in the innermost region  
of the disk will be enhanced. 

\subsection{Thermal Equilibrium Curves of low-$\beta$ Disks}

Numerical results revealed that as the cooling instability grows, 
the disk approaches to a quasi-steady state when a low-$\beta$ 
disk is formed. 
The cooling time scale is 
$t_{\rm cool} = 
\langle P \rangle / \langle Q_{\rm rad}^{-} \rangle \sim 160$, 
here $\langle ~ \rangle$ represents the average of a quantity 
over a range of $0 < z < 2$. 
Figure \ref{fig:fig3}b shows that the equatorial temperature stays 
$T \sim 10^{-5}$ at least for time scale of 
$t \sim 6000  \sim 20 t_{\rm dyn}$. 
This indicates that in this state,
heating balances with cooling. 
Since the dissipative heating rate of turbulent accretion disks 
is comparable to $ - \langle B_{\varpi} B_{\varphi}/4 \pi \rangle 
 \varpi (d \Omega/d \varpi)$
\citep{hir05}, the heating time scale can be estimated by 
$t_{\rm heat} \sim \langle P_{\rm gas} \rangle 
/[\langle B_{\varpi} B_{\varphi}
/4 \pi \rangle (3/2) \Omega]$, 
where $\Omega \propto \varpi^{-3/2}$. 
Before the thermal instability grows, since the disk is 
dominated by gas pressure, and $\langle B_{\varpi} B_{\varphi}
/ 4 \pi \rangle \sim 0.1 \langle P_{\rm gas} \rangle $ 
(see figure \ref{fig:bxby}a), 
$t_{\rm heat} = 10 /[(3/2) \Omega] \sim 2000$. 
After the cooling instability grows, the disk evolved toward a 
magnetic pressure dominated state in which $\langle B_{\varpi} 
B_{\varphi} / 4 \pi \rangle \sim 0.1 
\langle P_{\rm mag} \rangle$. 
Thus, the heating time scale at this stage is 
$t_{\rm heat} \sim \langle P_{\rm gas} \rangle 
/ [0.1 P_{\rm mag} \times (3/2) \Omega] 
\sim 1/ [(3/2) \Omega] \sim 200$ at $\varpi=35$. 
This time scale is comparable to the cooling time scale.

Numerical results indicate that steady equilibrium solutions of 
optically thin, low-$\beta$ disks exist. 
Thermal equilibrium curves of accretion disks can be obtained by 
equating the heating rate $Q^{+}$ and radiative cooling rate 
$Q_{\rm rad}^-$ or advective cooling rate $Q_{\rm adv}^-$, 
assuming $\alpha$-viscosity 
(\cite{abr95}; see also \cite{kat98}). 
When magnetic pressure is dominant, 
the vertically integrated pressure can be evaluated by 
$W \sim 2 B^2 H / (8 \pi)$, 
where $H$ is the half thickness of the disk. 
The vertical hydrostatic balance requires 
$W/(\Sigma H) \sim \Omega^2 H$. 
Based on the numerical results, we assume 
$\Phi_0 \equiv BH \sim {\it constant}$ at a fixed radius, 
where $\Phi_0$ is the azimuthal magnetic flux per unit radius. 
Thus $W \sim (BH)^2/(4 \pi H) \sim \Phi_0^2/(4 \pi H) $. 
By equating this equation with $W \sim \Sigma \Omega^2 H^2$, 
we obtain $ H \sim [\Phi_0^2/(4 \pi \Omega^2)]^{1/3} \Sigma^{-1/3}$.  
So we obtain the heating rate 
\begin{equation}
Q^{+} \sim \frac{3}{2} \alpha W \Omega 
      \sim \frac{3}{2} \alpha \Omega^{5/3} 
           \left( \frac{\Phi_0^2}{4 \pi} \right)^{2/3} \Sigma^{1/3}
      \propto \Sigma^{1/3} ~ .
\label{eqn:12}
\end{equation}
Since $Q^{+} \propto \dot{M}$, $\dot{M} \propto \Sigma^{1/3}$.

The radiative cooling rate $Q_{\rm rad}^{-} = Q^{'}_{\rm b} \rho^2 T^{1/2} 
(2H) = Q^{'}_{\rm b} \Sigma^2 T^{1/2}/(2H)$ 
is expressed as  
\begin{equation}
 Q_{\rm rad}^{-} = \frac{Q_{\rm b}^{'} }{2} 
                   \left( \frac{\Phi_0^2}{4 \pi} \right)^{-1/3}
                   \Omega^{2/3} \Sigma^{7/3} T^{1/2}
      \propto \Sigma^{7/3} T^{1/2} ~ .  
\label{eqn:13}
\end{equation}
By equating $Q^+$ and $Q_{\rm rad}^{-}$, 
we obtain 
\begin{equation}
T =   \left(
  \frac{ 3 \alpha \Omega \Phi_0^2 }{4 \pi Q^{'}_{\rm b}}
  \right)^2 \Sigma^{-4} ~ .
\label{eqn:14}
\end{equation}
When we adopt $\Phi_0^2 = (BH)^2 \sim 0.2$ and $\alpha \sim 0.1$ 
according to the results of our numerical simulation at 
$t=24100$ and $\varpi = 35$, we obtain
\begin{equation}
T \sim 0.007 \times \Sigma^{-4} ~ . 
\label{eqn:15}
\end{equation}

Figure \ref{fig:f9}a and \ref{fig:f9}b show the thermal equilibrium 
curves of accretion disks. 
Thin curves show the equilibrium curves of conventional accretion 
disks (\cite{abr95}; see also \cite{kat98}). 
Thick solid curves show the optically thin low-$\beta$ branch 
$T \propto \Sigma^{-4}$ depicted by using equation (\ref{eqn:15}) 
and $\dot{M} \propto \Sigma^{1/3}$.  
These curves should be connected to the optically thin, gas pressure 
dominated branch as indicated by the thick dotted curves. 
The connection between 
the hot gas pressure dominated branch and the magnetic pressure 
dominated cool branch should be studied including the advective 
cooling and will be presented in a separate paper 
(Oda et al. 2006 in preparation). 

Let us discuss the thermal stability of the low-$\beta$ branch. 
Since $Q^+ \propto \Sigma^{1/3}$ (equation (\ref{eqn:12})) 
and $Q_{\rm rad}^{-} \propto 
\Sigma^{7/3} T^{1/2}$ (equation (\ref{eqn:13})), 
$(\partial Q^+ / \partial T )_{\Sigma} = 0$ 
and $(\partial Q_{\rm rad}^{-} / \partial T)_{\Sigma} 
\propto  Q_{\rm rad}^{-}/(2 T) > 0 $, 
\begin{equation}
\left( \frac{\partial \ln Q_{\rm rad}^{-}}{\partial \ln T}
\right)_{\Sigma} 
> \left( \frac{\partial \ln Q^+}{\partial \ln T}
\right)_{\Sigma} ~ ~. 
\end{equation}
This indicates that 
the low-$\beta$ branch is thermally stable (e.g., \cite{pri81}). 
The low-$\beta$ branch is also secularly stable because 
$\dot{M} \propto \Sigma^{1/3}$ at a fixed radius, thus the 
criterion for the secular stability (e.g., \cite{pri81}),  
\begin{equation}
\left( \frac{\partial \dot{M}}{\partial \Sigma} 
 \right)_{Q^+=Q_{\rm rad}^-} > 0 
\end{equation}
is satisfied.

Figure \ref{fig:f9}c shows the relation between $\Sigma$ and $T$ 
obtained from our numerical simulation.
Here, $T$ is the equatorial temperature. 
The symbols show the surface density and mass accretion rate at 
$\varpi = 35$ averaged in azimuthal direction and 
integrated in $|z| < 10$. 
The time range plotted is $ 21000 < t < 35000$. 
An arrow indicates the time ($t = 24100$) when we included the 
radiative cooling. 
The dashed line shows the relation $T \propto \Sigma^{-4}$  
expected from the above theory. 
Figure \ref{fig:f9}d shows the relation between $\Sigma$ and $\dot{M}$.  
As the disk approaches the quasi-steady low-$\beta$ state, 
it follows the theoretically expected relation 
$\dot{M} \propto \Sigma^{1/3}$. 

\subsection{Dependence on Initial Magnetic Field Configuration}

We would like to discuss the dependence of numerical results on 
the initial configuration of magnetic fields. 
Since the growth rate of MRI is larger when the initial magnetic 
field has poloidal component than the case with purely toroidal 
magnetic field (e.g., \cite{haw01}; \cite{kat04}), 
the accretion from the initial torus proceeds 
faster in models with initially poloidal magnetic field. 
In the nonlinear stage, however, the structure and 
strength of magnetic fields 
do not depend significantly on the initial configuration of 
magnetic fields so long as 
the initial magnetic field is weak and confined inside the disk. 
In simulations starting from weak poloidal magnetic 
fields embedded inside the disk  
(e.g., \cite{haw01}; \cite{kat04}), 
the amplification of magnetic field saturates when 
$\beta \sim 10$. 
The magnetic field strength at this stage and the ratio 
of poloidal magnetic field to toroidal magnetic field 
are comparable to 
that for purely azimuthal magnetic fields. 
Our simulation results also show significant mass outflows 
in the coronal region, consistent with the results starting 
from weak poloidal magnetic fields. 
Thus, when radiative cooling is included after a 
magnetically turbulent accretion disk is formed, 
subsequent evolution of the disk will not be sensitive 
to the initial configuration of magnetic fields.

\subsection{Comparison with Observations}

\begin{figure}
 \epsscale{1.0}
 \plotone{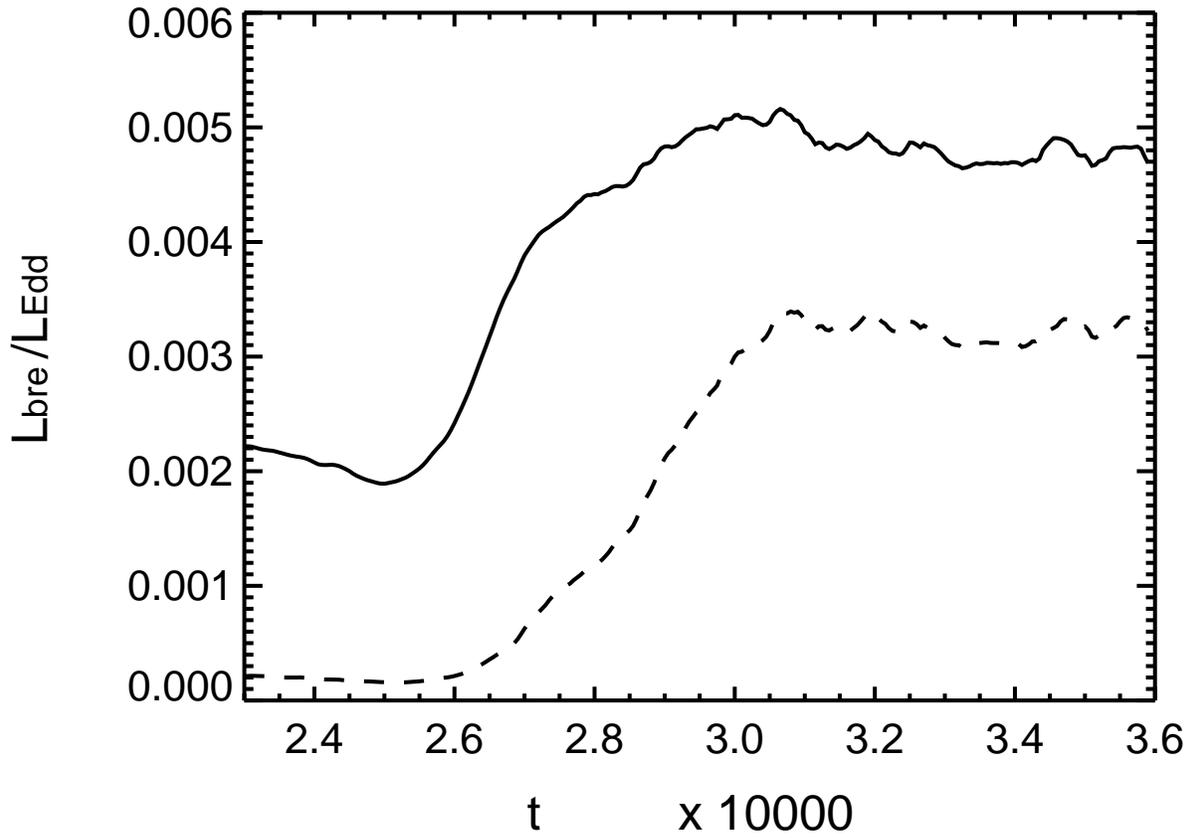}
 \caption{
Time evolution of the bremsstrahlung luminosity integrated in 
$ 4 < \varpi < 50$, $ |z| < 30$ (solid curve) and 
in $4 < \varpi < 50$, $|z| < 1$ (dashed curve). }
 \label{fig:lum}
\end{figure}

\begin{figure}
 \epsscale{1.0}
 \plotone{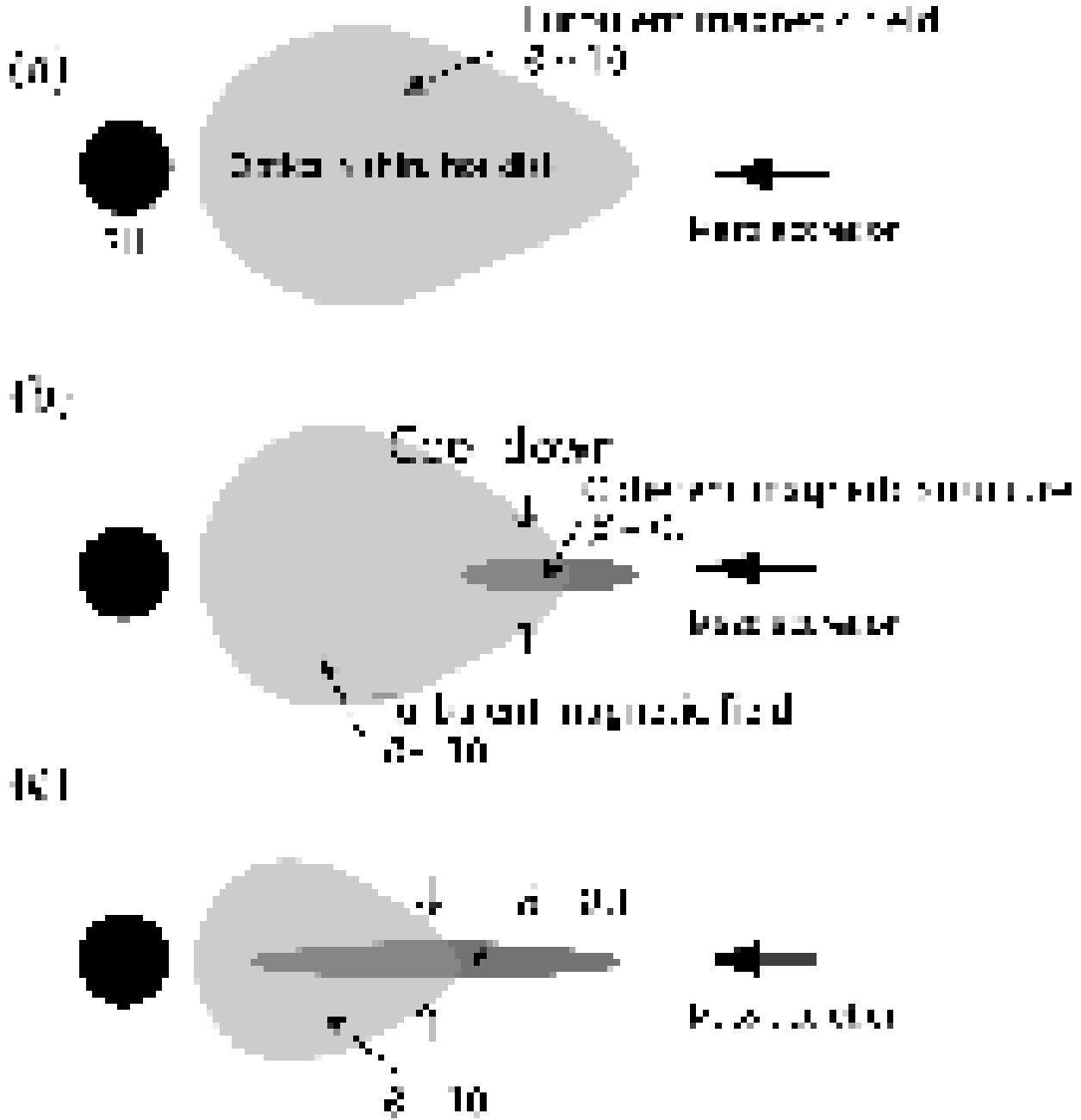}
  \caption{
A schematic picture of state transitions in black hole candidates. 
(a) Low/hard state, 
(b) Onset of the cooling instability, 
(c) The transition radius between the hot disk and cool low-$\beta$ 
disk moves inward as accretion rate increases.  
}
  \label{fig:f8}
\end{figure}

Let us compare our numerical results with observations of 
black hole transients. 
\citet{hom05} 
pointed out that during the transition from the low/hard state 
to the high/soft state, 
GX 339-4 shows a sub-transition from hard intermediate state 
dominated by power-law X-ray radiation to soft intermediated state 
dominated by radiation from optically thick disk. 
The X-ray luminosity of the hard intermediate state is 1-10\% of the 
Eddington luminosity in GX339-4 \citep{rem05}. 
This luminosity is much higher than the critical luminosity 
above which the optically thin, hot gas pressure dominated disk 
becomes thermally unstable. 
In our simulations, the critical luminosity for this transition 
is about 0.2\% of the Eddington luminosity 
(see figure \ref{fig:lum}). 
The magnetic pressure dominated disk shows hard X-ray spectrum 
because it is still optically thin. 
Thus, numerical results suggest that 
the luminous disk in the luminous hard state stays in the 
thermally stable low-$\beta$ branch. 

\begin{figure}
  \epsscale{1.0}
  \plotone{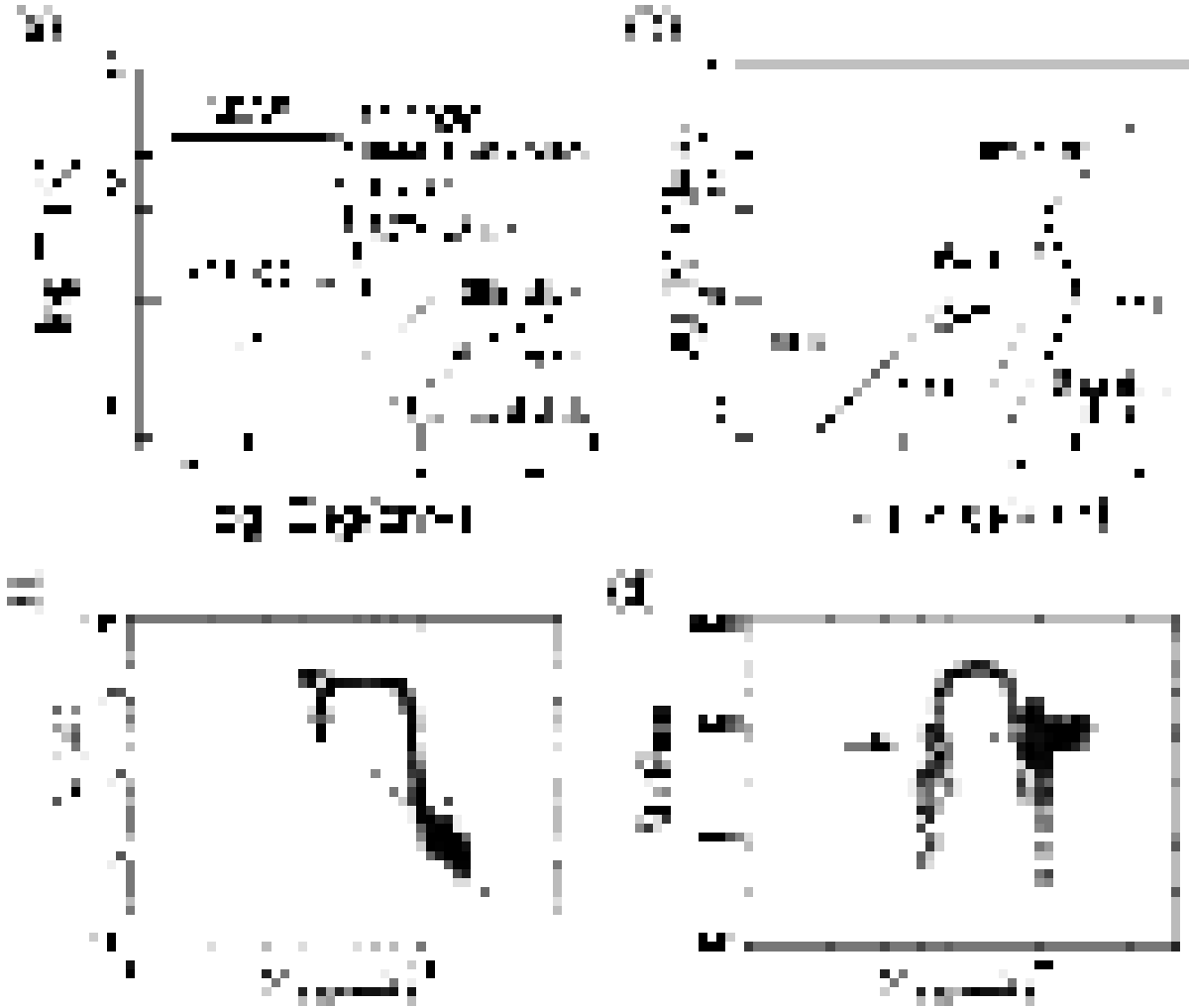}
  \caption{
(a) Thermal equilibrium curves of accretion disks (solid curves). 
The thick line shows the optically thin, low-$\beta$ branch 
which is thermally stable. 
(b) Schematic picture of thermal equilibrium curves in 
the $\Sigma-\dot{M}$ plane. 
The thick line shows the optically thin, low-$\beta$ branch 
which is secularly stable. 
(c) The relation between surface density and equatorial temperature 
at $\varpi=35$. 
The dashed line shows the theoretically obtained relation of the 
the low-$\beta$ disk $T \propto \Sigma^{-4}$. 
An arrow shows the point at $t=24100$ when radiative 
cooling term is included. 
(d) The relation of surface density and mass accretion rate 
at $\varpi = 35$. 
The dashed line shows the theoretically expected curve for 
low-$\beta$ disk 
$ \dot{M} \propto \Sigma^{1/3}$. }
  \label{fig:f9}
\end{figure}

The X-ray luminosity in our simulation results saturated at 
about 0.5\% of the Eddington luminosity because the mass 
accretion rate from the outer torus saturated at about 
$\dot{M} \sim \dot{M}_{\rm Edd}$ 
(figure \ref{fig:f9}b). 
We expect that when the mass accretion rate from the outer region 
continues to increase, the transition radius between the 
inner hot flow and the outer cool low-$\beta$ disk will 
move inward and the X-ray luminosity will continue to 
increase. 
Such simulations will be possible by starting simulations 
from a torus initially located at larger radius. 

Since the low-$\beta$ disk is supported by magnetic pressure,
the transition to the optically thick (soft intermediate) state 
needs the dissipation or expulsion of magnetic fields 
supporting the disk. 
Recently, \citet{fen04} pointed out that in galactic black hole 
candidates, relativistically moving blobs are ejected 
during the transition from a low/hard state to a high/soft state. 
Our numerical results suggest that such ejections originate from 
the releases of magnetic energy stored in the low-$\beta$ disk. 
When magnetic flux inside the disk is ejected by such 
explosive events, 
the disk will again be supported by gas pressure and 
will be able to shrink further in the vertical direction. 
Extraction of magnetic flux from the low-$\beta$ disk 
enables the disk to complete the transition to the optically thick, 
high/soft state. 

Synchrotron and/or Compton scattering are not included in 
the present study. 
If included, magnetically supported disks are expected to 
emit enhanced radiation from the inner parts of the disk.  
It may resolve the issue of overproduction of 
bremsstrahlung from the outer parts of the low luminosity MHD flow 
(see \cite{ohs05} for discussion).

In this paper, we assumed single temperature plasma. 
In optically thin, hot accretion disks, 
electron temperature can be lower than the ion temperature. 
Such two temperature global disk models including radial advection 
were studied by Nakamura et al. (1997; see 
also \cite{man97}). 
To fit the observational spectra, two-temperature nature of 
accretion flow should be considered in the simulations. 
Works are now in progress to extend our simulation code to 
two temperature plasmas.

\bigskip

We are grateful to S. Mineshige, R.A. Remillard, M.A. Abramowicz, 
D. Meier, S. Hirose and Y. Kato for discussion. 
Part of this work was carried out when M.M. and R.M. attended the 
KITP program on ''Physics of Astrophysical Outflows and 
Accretion Disks'' at UCSB. 
Numerical computations were carried out by VPP5000 at NAOJ 
(P.I. MM). 
This work is supported in part by Japan Society for the 
Promotion of Science (JSPS) Research Fellowships for 
Young Scientists (MM: 16-1907, 17-1907),  
Grant-in-Aid for Scientific Research of the Ministry of Education, 
Culture, Sports, Science and Technology 
(RM: 15037202, 16340052, 17030003), 
ACT-JST of Japan Science and Technology corporation, 
and by the National Science Foundation under Grant No. PHY99-07949.

\newpage


\newpage

\begin{table}
 \caption{Units adopted in this paper.}
 \label{tab1}
\begin{center}
\begin{tabular}{lll} \hline
& Numerical Unit & Physical Unit \\ 
\hline 
length & $r_g$ &  $3.0 \times 10^6 (M/10M_{\odot})$ cm \\
velocity & $c$ & $ 2.9979 \times 10^{10}$ cm ${\rm s}^{-1}$ \\
time & $t_0$ & $ 10^{-4} (M/10M_{\odot})$ sec \\
density  & $ \rho_0 $ & $8.3 \times 10^{-7} (M/10 M_{\odot})$ 
g  ${\rm cm}^{-3}$ \\
temperature & $T_0$ & $1.1 \times 10^{13}$ K \\
luminosity & $L_{\rm Edd}$ 
& $1.25 \times 10^{39} (M/10M_{\odot})$ erg ${\rm s}^{-1}$ \\ \hline
\end{tabular}
\end{center}
\end{table}


\end{document}